# The critical factors affecting E-Government adoption: A Conceptual Framework in Vietnam

NGO TAN VU KHANH

School of IT Business – SOOGSIL University – Seoul - South Korea;
ngotanvukhanh@gmail.com

**Abstract**

Electronic government (e-government) has established as an effective mechanism for increasing government productivity and efficiency and a key enabler of citizen-centric services. However, e-government implementation is surrounded by technological, governing and social issues, which have to be considered and treated carefully in order to facilitate this change. This research attempts to explore and investigate the key challenges that influence e-government implementation and the factors influencing citizen adoption in Vietnam. It develops a conceptual framework on the basis of existing experiences drawn from administrative reforms. Survey data from public employee will be used to test the proposed hypothesis and the model. Therefore, this research has identified factors that determine if the citizen will adopt E-government services and thereby aiding governments in accessing what is required to increase adoption. We will also highlight several research, practitioner and policy implications.

## 1. Introduction

Successful diffusion of information communications technology (ICT) has technologies triggered the usage of Internet, e-commerce, and eventually in electronic government (e-government). Tat-Kei Ho (2002) explains that explosive growth in the Internet usage and rapid development of e-commerce in private sector have put growing pressure on the public sector to serve citizens electronically. Electronic government initiatives have often sounded very promising but have been difficult to implement. The challenge lies in the implementation of e-government successfully. One likely barrier is that e-government is approached as if in a universal context, which can be generalized across the globe. On the other hand, the individual countries' contextual imperatives, culture and conditions vary. Therefore, a universal approach is less likely to be effective in all contextual settings. This research mainly focuses on e-government in Vietnam, explore important factors affecting the adoption and implementation of e-government in Vietnam.

ICT has been being used in the Vietnam public sector for more than 22 years and the advent of the Internet has given this usage more than just a new name-e- Government-and a higher profile. During last 22 years, 5 big projects have been implemented: two of which were financially supported by the French Government (in the 1991-1993 and

1994-1996 periods); one invested by the State budget (a part of the national IT program, period 1996-1998); and the other under the Prime Minister's Decision in 1997. However, the achievements were still very limited. Going along with the new trend of e-Government in the world, in 2001 the Vietnam government decided to start the new project (Project 112) which was considered as the milestone for e-Government in Vietnam. Unfortunately, in April 2007 Project 112 was halted. Start the year 2007, a grand scheme named "ICT development in Vietnam "by the Ministry of Information and Communication implementation (2008-2013) and the results are not as expected. Indeed, this phenomenon has not only occurred in Vietnam. The chief information expert of the World Bank indicated that "among the e-Government projects in developing countries, according to estimation, 35 totally failed, and 50 partially failed. Only 15 can be considered completely successful"(Hu et al., 2005). Obviously the benefits of e-Government (i.e. transference, efficiency, participant as defined by OECD,2003) could not be delivered to its stakeholders if the failure occurred. Further, failures come at a higher price for the world's poorer countries since their resources are limited and the wasteful capital should be invested in the other profitable projects. The research has explored the area of e-government adoption. it identified the factor influence on e-government. In addition, it will seek to understand how literature identified the factor in e-government in order to compare the propose and practical application in the real world setting. The research will also determine the obstacles and benefits in adopting this concept .In the view of the above points, this contribute to both, theoretical as well as practical literature, there by study will justifying its worth by bridging the gap in the extant literature about the e-government adoption. And this is actually a specific study to save costs for e-government projects success in the future.

As a result the paper sought to answer three key research questions. (i) What are the critical factors of e-government delivery in terms of implementation? (ii) What is the level of validity of the proposed factors in Vietnam in terms of e-government implementation? (iii) What implications may emerge in implementing e-government in the context of Vietnam? An investigation into the three research questions in the current study would help to identify the most influential factor of e-government adoption. This paper is organized as follows. In the next section, we briefly introduce the context of ICT in Vietnam and the e-government development in Vietnam. Then, theoretical background and the research model are presented, which is followed by a detailed report on the results of the study. Thereafter, results are discussed with a number of implications and conclusions. Finally, limitations of this study and implications for future studies are discussed as well.

## 2. Related research

### 2.1. Context of ICT in Vietnam

At the national level, government in Vietnam consists of the dual structures of the Communist Party and the National Government. The country is organized at local government level into 64 provinces. The provinces are divided into 588 districts, which are further subdivided into 9069 communes (General Statistics Office of Vietnam 2013). According to the CIA World Factbook (2013) the area of Vietnam is

329,560 sq km. Total population is 90 million (November, 2013.) with a median age of 26.4 years, comprising 26.3% of 0-14 years, 67.9% of 15-64 years, and only 5.8% of 65 years and over (Central Intelligence Agency 2013). Approximately 60 million live in a rural setting with farming as the main occupation (General Statistics Office of Vietnam 2009).

Vietnam has undertaken a far-reaching process of economic reform known as 'Doimoi' since 1986. The Government of Vietnam committed to increased economic liberalization and enacted structural reforms needed to modernize the economy and to produce more competitive, export-driven industries. Achievements of "Doimoi" have been spectacular. According to a World Bank (2003) report, the progress made in Vietnam in alleviating poverty has been one of the greatest success stories in world economic development in recent years. Poverty rates measured at international levels have halved from 58% in 1992 to 37% in 1997 and 29% in 2002 (World Bank 2003) and further down to 24.1% in 2004 (ADB 2005). Together with poverty reduction, the country had very successful economic growth since 1990 with around 8% annual GDP growth from 1990 to 1997, 5.5% from 1998 to 2000, over 7% from 2002 to 2005 and 8.4% in 2006, making it the world's second-fastest growing economy. The country also achieved a high rate of increase in exports, from US$9.1 billion in 2001 to US$16.5 billion in 2002 or over 12% a year (Dapice and Fellow 2003). In 2013, nominal GDP and per capita GDP were US$176 billion and US$1.960, respectively (General Statistics Office of Vietnam 2013).

It has been more than one decade since the Internet started to have been used in Vietnam. Vietnam connected the world in 2000, the Internet users was a small figures, just 0.3% of the population in 2000. However, the Internet is growing fast, much faster than in any other Asian countries in 2011. Over the last ten years 2000-2010, Internet usage has grown by 12.4 times in Vietnam. This is the highest level of penetration in the Asian countries. After five years from 2000, this number was up to 12.8%; and 17.9% in 2007; 24.0% in 2008; and 25.7% of Vietnam population in 2009. At the present, according to the Vietnam Internet Network Information Center, Vietnam ranks 18/20 countries with the largest number of Internet users in the world, ranking eighth Asia and ranks third in Southeast Asia with 31,302,752 Internet users as of Dec 2013, 35.53% of the population. The number of Vietnam Internet user has increased more than 15 times compare with that of 2001. Such an advantage of Internet is very potential for development of e-government service.

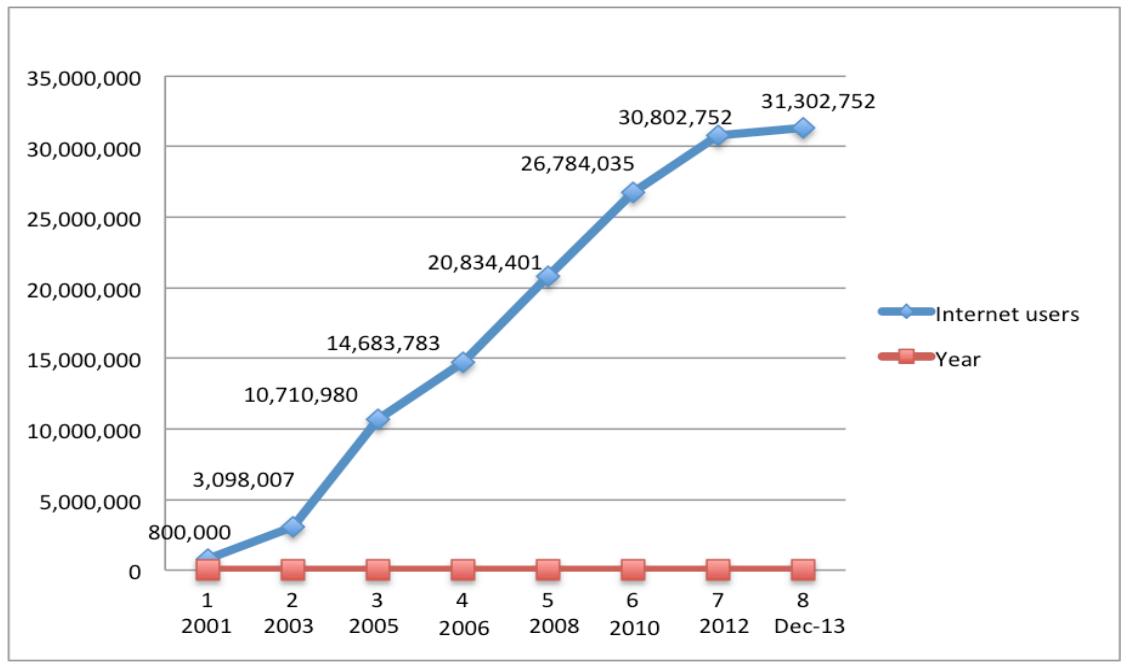

Figure 1: Vietnam Internet users

*Source: Vietnam Internet Network Information center (VNNIC)*

Amongst small enterprises in Vietnam the diffusion of ICTs and the Internet in particular has been slow. In a case study in 2012 of nine traditional villages focusing on small enterprises and e- commerce, research identified the following difficulties: unreliable technological infrastructure, lack of legal infrastructure, blocking of Internet access due to security concerns. However, despite the infrastructural constraints and difficult conditions, a number of small enterprises are engaging in e-business. E-government cannot be achieved without the availability of telecommunications infrastructure. With the increase has come an 'explosion' in the development of the Vietnamese ICT sector since the turn of the millennium. According to statistics presented by the World Bank (World Bank and WITSA 2009), Vietnam's investment in ICT infrastructure as a percentage of GDP was of a similar order to that of larger economies in the region: Vietnam spent 5.51% of GDP (4.85% of GDP 2008), while South Korea spent 9.07%(9.1% of GDP 2011); Malaysia 9.7%; Hongkong 9.2% Singapore 8.4% and Thailand 6.2%. Clearly the absolute figure spent in Vietnam was much lower than that of these countries, given the differentials in GDP terms. Nevertheless the level of commitment being made in percentage terms sets a good base for long-term development.

Vietnam ICT Summit 2013 reported that the Vietnamese government was enthusiastic in encouraging e-commerce and e-government projects, introducing ICT strategies and ICT parks to attract foreign companies. Policy makers have also been keen to foster B2B e-commerce to promote export industries, however both B2B and B2C e-commerce have been described as negligible. Vietnam ICT Summit 2013 identified some challenges to ICT expansion in Vietnam, namely: lack of sufficient competition in the ICT sector; high piracy rates; and a shortage of ICT skilled labor.

**2.2. E-government development in Vietnam**

Vietnam Governments are embarking on providing e-access to their citizens for the public service delivery. It is on the agenda of many governments to benefit form the emerging shift from the conventional methods of paper work to the digital age. This agenda is triggered by the efficiencies and effectiveness attributed to the e-government perspectives. Such attributes have dual purpose: cost and time efficiencies from the supplier's perspectives and the satisfaction of recipients of the quality services. Therefore, enhancing citizens' satisfaction through quality service delivery in terms of time and space is the paramount concern in this direction.

Year 2010 is considered a milestone in the deployment of e-Government (e-Government) in Vietnam, was the end of the year of IT application in State agencies activities that the Government set out in the period 2006 - 2010. At that year also driven IT solutions period 2011 - 2015 with the implementation of Decision 43/2008/QD-TTg and 48/2009/QD-TTg of ICT application in State agencies plans national "ICT Application in the State agency period 2011 - 2015" with a total investment of 1.700 billion. In addition, there are 20 national projects on IT application has been approved by the Prime Minister. Various policies above showed that the government of Vietnam has a political will for successful implementation of e-government in Vietnam, even to the level of local government. Moreover, with implementation of the Act Number 43 in 2010 regarding Information and Electronic Transactions (ITE). This Act supported to the public service transactions through e-government. However the completeness of the policy is yet to yield significant results for the development of e-government in Vietnam. It can be proved by E-Government Readiness Rank, according to the United Nation, Vietnam is still in low rank among other countries in Southeast Asia, moreover for Global Rank of E-Government Readiness (United Nations, 2010). Vietnam's position in the E-Government Readiness in both Southeast Asia and global can be explained in the following table 1

| | Country | Global Rank | | | | |
|---|---|---|---|---|---|---|
| | | 2004 | 2005 | 2008 | 2010 | 2012 |
| 1 | Singapore | 8 | 7 | 23 | 11 | 10 |
| 2 | Malaysia | 43 | 42 | 34 | 32 | 40 |
| 3 | Brunei Darussalam | 63 | 73 | 87 | 68 | 54 |
| 4 | Vietnam | 112 | 105 | 91 | 90 | 83 |
| 5 | Philippines | 47 | 41 | 66 | 78 | 88 |
| 6 | Thailand | 50 | 50 | 64 | 76 | 92 |
| 7 | Indonesia | 85 | 86 | 106 | 109 | 97 |
| 8 | Lao | 147 | 147 | 156 | 151 | 153 |
| 9 | Cambodia | 129 | 128 | 139 | 140 | 155 |
| 10 | Myanmar | 123 | 129 | 144 | 141 | 160 |

| 11 | Timor Leste | 174 | 144 | 155 | 162 | 170 |

Table 1: E-Government Development in South Eastern Asia

Among Southeast Asian countries, Vietnam's position is in the fourth rank, increased 7 ranking compared to other countries in the world and increased 2 ranking compared with Southeast Asian. Lower than Malaysia and Singapore, and rank higher than 7 countries. As for the global rank, Vietnam's position has increased for each year. This facts are conformity to the number of Internet users in Vietnam, which has increased very significantly. In 2001 when the prime Minister Decision on e-government Vietnam was issued, Internet users are only 8 hundred thousand (Vietnam Internet Center,20013), whereas in 2010 the Vietnam Internet users increased dramatically to 26 million users and in 2012 growing to 30 million users (Vietnam Internet Center,2013). Based on these facts, a research regarding e-government adoption of Vietnam is very necessary to be done, since the success of e-government implementation is dependent not only government support, but also on citizen's willingness to accept and adopt those e-government service

## 3. Theoretical backgrounds and the research model

### 3.1. E-government processes

A number of research papers in e-Government were published in few years ago to help Researchers to improve government responsiveness, service quality,, accessibility and convenience to both citizens living in urban and rural area. Their effort can be categorized into few issues: (i) The concept, theories, policy, structure, initiatives, history, impact, key principles, challenges, and development success factors of e-Government; (ii) knowledge spillover, the technology application, , innovative efforts and approach to facilitate e-Government implementation and evaluation; (iii) management support or implementation strategies such as framework for managing the lifecycle of transactional e-Government services to facilitating the e-Government services. To provide more intuitive and maintainable lifecycle for electronic tax submission (one of the e-government service to citizens), government must overcome shortcoming happen during the lifecycle such as implicit knowledge, code reusability, user interaction, communication with back-end system, business reengineering required to upgrade to workflow flexibility and resolve security issue (Vassilakis, Laskaridis, Lepouras, Rouvas, & Georgiadis, 2003) (iv) the assessment, measurement of e-Government services provided to public sector and its effects on economic, social benefits of the implementation; (v) key factors affecting acceptance, expectation and usage intention of e-Government services. The study on "Utilization of e-government services" discussed that compatibility; trustworthiness and perceived ease of use have direct positive relationship towards citizens' intention to use the e-Government services (Carter & Belanger, 2005).

### 3.2. Governing Factors

#### 3.2.1. Vision

Vision has become the buzzwords from the 1990s. "Vision" exists from the top multinational right down to the comer shop. According to Wills (1994), a vision can exert a strong pulling force which helps to keep the organization aligned – depend on emotional aspect to describe. Based on Allen (1995), a vision must:(i) Be coherent enough to create a recognizable picture of the future; (ii) Be powerful enough to generate commitment to performance; (ii) Emphasize what realistically can be and Clarify what should be.

Establishing a broad vision of E-government should start from the planning process which is the shared gold of the society. The board vision sequent the large goals and concerns of a society (PCEP, 2002). Citizens should be built in the E-government vision. Moreover, the demand side of business should also under the E-government vision. Government support and assistance should again be available at anywhere and anytime to aid businesses succeed, and regulate the environment for a healthy electronic market economy (Stamoulis and Georgiadis, 2000). A vision is a commitment to establishing rethinking and reviewing who we are and what we are here to do (Allen, 1995). Hitchcock (1996) showed Rules for the vision should be framed:(i) It must be phrased positively; it should be stated in terms of moving towards something 'that is wanted; (ii) It should be as specific as possible; preferably, the evidence should be in a form that is sensory-based, that is, what could be seen, heard or felt; (iii) It must be possible to have an input into the achievement of the vision, to have ownership. Hence, the authors of this research hypothesize that Vision affects E-Government Adoption positively.

*H1: Vision will have a positive effect to Critical Factors for E-Government Adoption.*

### 3.2.2. Top management support

Widespread organizational support and acceptance from the top management are the succeed key for apply a new way of doing business or of a new technology. According to top management is defined as not only the president and CEO, but also all managers who have the authority to establish and enforce policies and guidelines (Cavaness and Manoochehri, 1993). The top management support, Holland, Light and Gibson (1999) showed it as the positive commitment, enthusiasm and support of senior management for the project. The top management is the permission of the person, or group of people, who directs and controls the highest level of organization. Top management support is necessary during the implementation. The project must receive approval from top management (Bingi, 1999; Buckhout, 1999; Sumner, 1999), and align with strategic business goals (Sumner, 1999). This can be achieved by tying management bonuses to project success (Wee, 2000).

Top management support is one of the important and critical success factor for E-government adoption. Top management needs to publicly and explicitly identify the project as a top priority (Wee, 2000). Senior management must be committed with its own involvement and willingness to allocate valuable resources to the implementation effort (Holland et al., 1999). This involves providing the needed people for the implementation and giving an appropriate amount of time to get the job done (Roberts and Barrar, 1992). The top management support is needed for E-government initiatives. There are many political decisions from top management during the E-government developing period. Governments need to define objectives in realizing their E-government system. The result shows that a high degree of top management support plays a significant role on E-Government Adoption.

*H2: Top management support will have a positive effect to Critical Factors for E-Government Adoption.*

### 3.2.3. Leadership:

Basing on Zairi (1994), "Nowadays leadership is considered as a must for survival. It comes from the level of inspiration, commitment generated and corporate extermination to perform". Organizations and researchers have been obsessed over the last four decades with leadership (Kets De Vries, 1993; Goffee and Jones, 2000; Conger and Toegel, 2002, Higgs et al., 2003). Leadership has been described as a process, but most theories and research on leadership look at a person to gain understanding (Bernard, 1926; Blake, Shepard and Mouton, 1964; Fiedler, 1967; House and Mitchell, 1974; Drath and Palus, 1994).

According to Morden (1996), leadership means getting results through people. Leadership is defined by the behaviors, traits and qualities of a leader. Basing on Klagge (1996), leadership refers to a phenomenon similar to trailblazing, where individuals are out in front of others exploring virgin territories, mapping new pathways, and setting the pace. E-government adoption comes more popular and more important for every country around the world. In order to adopt and applying the E-government system, there are some important issues to be considered. Leadership is one of the core for success, starting with the definition. Leaders of government need to know what is limited sense of E-government, through moving government E-services, miss larger opportunities in the future.

Leadership is one of the important factors for the E-govenment success hinges. A critical pre-condition in e-govenment adoption is a strong leadership with vision Heeks, 2002). Moreover, Ke and Wei (2004) indicated that strong leadership with vision is a crucial factor for e-government success.

*H3: Leadership will have a positive effect to Critical Factors for E-Government Adoption.*

### 3.2.4. Funding:

According to Okiy (2005), "The importance of funding in providing excellent service cannot be over emphasized. It is the glue that holds the building, collections and staff together and allows attaining goals". Clearly, funding is the factor which promote the success of E-government. Fund in governmental accounting is defined: "representing a distinct phase of the activities of goverranent and is controlled by a self-balancing group of accounts in which all of the financial transactions of the particular phase are recorded and the fund is both a sum of resources and an independent accounting entity" (http://www.michigan.gov).

Lack of funding in a project is certainly a disincentive, especially when adopting an innovation means that individuals must go through a learning curve and take on new responsibilities as a result of developing expertise (Sherry, 2003). Financial savings to governments through alpplying E-service will occur just from the medium-to-long term. Initial start-up costs will be high, in the short term, especially for parallel manual E-government system for any length of time. E-government is mainly related to lack of funding (Akomode et al., 2002). In the US, lacking of financial resources as a barrier to applying E-government for over half (57.1%) of city and county governments (ICMA, 2002). Funding was as the greatest obstacle to moving county government services to the Online services by 70% of the respondents (NACO, 2000). Hence, the authors of this research hypothesize that Funding affects E-Government Adoption positively.

*H4: Funding will have a positive effect to Critical Factors for E-Government Adoption.*

### 3.3. Social Factor

### 3.3.1. Social Influence Statements

Public organizations that have introduced electronic services have done so by radically transforming their organizational structures using the latest innovations in technology (Bjorck, 2004). Barley (1986:79) defines social structure as patterned action, interaction, behaviours, and awareness. Davidson and Chismar (2007:741) posit that social structures often become a taken for granted aspect of social life. Given this context, institutional theory explains how external pressures can control the input processes of an organization, resulting in actions that would ultimately increase the quality of services and improve customer satisfaction. By gaining customers' satisfaction, organizations will be able to reduce any negative external pressure of social behaviour within an e-government context. This procedure will help public organizations to transform from traditional services to online services with respect to social behaviour. Worthy of note here is the result that transformation will have on the implementation of new rules, procedures and organizational processes that relate to and shape social behaviours (Liang et al.,2007; Teo et al., 2003).

The social dimensions of e-government related institutional change have been identified by a number of scholars such as Kim et al., (2009); Al-Gahtani et al., (2007); and Yildiz, (2007). In addition, Heeks (2005; 53) argues that e- government is connected to the social context in which it is deployed. This can be seen firstly in the way that technology can impact that social context. Moreover, Liao and Jeng, (2005: 505) argue that Public administration involves planning and implementing various policies in order to solve various complex problems posed by the social, political, and economic environment. Therefore, significant social, organizational and technical challenges will need to be understood well and to prevail over those efforts that attempt to accomplish governmental transformation (Affisco and Soliman, 2006). This is particularly significant as what looks like technologically determinist research to one person might look like socially determinist research to another (Heeks and Bailur, 2007: 245). Therefore, Social Influence Statements is hypothesized to have an indirect impact on E-Government Adoption.

*H6: Social Influence Statements will have a positive effect to Critical Factors for E-Government Adoption.*

### 3.3.2. Awareness:

Basing on Dourish and Bellotti (1992), awareness is "an understanding of the activities of others, which provides a context for your own activity". Awareness includes using the mass media to introduce the concept of E-givernment system for people in the public sectors, conducting seminars or workshops to encourage the public sectors' work force to apply the concepts as their daily operations. A package of activities could be delivered that includes (Heeks, 2001b) seminars and training workshops, web-based documentation, individual meetings, and support for monitoring and project evaluation.

According to Papazafeiropoulou, Pouloudi and Doukidis (2002) magazines, articles, videotapes, websites, books, newsletters, brochures, CD-ROMs, presentations, road-maps, guidelines, and case studies are awareness material. The rapid growth of E-government technologies and practices has created a tremendous need for awareness creation in organisations which seem to lack the necessary information about technology (Papazafeiropoulou, Pouloudi and Doukidis, 2002). The Government agencies would realise the benefits of E-government system is the important of implementing E-government. Reducing the time and cost of providing services to empowering the employees, reducing bureaucracy, and increasing efficiency are these benefits.

Basing on Kotter (1996), individuals are willing to accept change (e. g. e-government initiatives) if the potential benefits are outlined and they believe that the transformation is possible. Developing economic of the country, inducing actual use of E-government system are the purpose of implementing appropriate public awareness programme which build on the favourable sentiments towards the E-government system. Hence, we propose the hypothesize:

*H6: Awareness will have a positive effect to Critical Factors for E-Government Adoption.*

### 3.3.3. Training:

Training and education are being given the importance that they hey deserve throughout industry and business. If people cannot use the new technologies, they cannot take responsibility for their own quality. Because the customer requirements are more and more stringent, so Skills can quickly unsatisfy. Training is important and necessary, but it is also costly (Read and Kleiner, 1996). Keep (1989) suggests that training can be viewed either as a cost or as an investment. According to Greig (1997), in general terms, the analysis divides in- service training into three classifications: (i) Training that can be conducted most readily and effectively outside the enterprise; (ii) Training that can be carried out most readily and effectively inside the enterprise; (iii) Training that, both technically and cost-effectively, can be delivered equally well outside or inside the enterprise.

Priority human capacities for E-government system are combined: those who have knowledge about technology, business and information in governance. Moreover, be hybridised into broader change agents is one of the requirement for IT professionals in E-government system who combine ICT and IS skills with knowledge of the public sector, civil society and change management. Public sector managers need to be hybridised towards a broader skill set that includes an understanding of information systems and ICTs (Heeks, 2001b).Training is adequate to meet and maintain e-government skills (AONSW, 2001). Also, the re-engineering of work processes needs to manage well, as well as retraining and educating the relevant staff members (Jupp, 2001). The importance of training, hands-on support, and a proactive stance towards adjusting the technology to the work have been identified as important both by practitioners (Keselica, 1994; Smith, 1996; Lloyd and Whitehead, 1996) and researchers (Bullen and Bennett, 1990; Rogers, 1994; Orlikowski et al., 1995, Karsten et al., 1997). Therefore, training is hypothesized to have an indirect impact on E-Government Adoption

*H7: Training will have a positive effect to Critical Factors for E-Government Adoption.*

### 3.4. Technology Factor

### 3.4.1. IT Infrastructure:

Infrastructure is one of the most important word not specific to IT. Infrastructure is probably most visible: we can see streets and other structures for transportation and logistics (Gray, 1998; Suomi, 2002), public buildings such as schools, museums and libraries (Coult, 2001; Hopkins, 2001). Infrastructure can be seen too in abstract things such as legislation, education system, public health care system, different markets and governance structures (Hypp6nen, Salmivalli and Suomi, 2005). The word infrastructure is also used in the area of IT (Broadbent, Weill and St Clair, 1999; Broadbent, Weill and Nco, 1999), and IT is more and more important role in public administration (Gore, 1993; Bellamy and Taylor, 1998; Heeks, 1999).

in the late 1980s before the Internet emerged, the government was already actively pursuing IT to improve operating efficiency and to enhance internal communication (Kraemer and King, 1977; King,. 1982; Fletcher et al., 1992; Norris and Kraemer, 1996; Brown, 1999). IT can help goverment public sectors to increase productivity and performance, improve policy-making, and provide better public services to the citizens (Akbulut, 2002). Moreover, there is an opportunity to derive productivity and business benefits from an intelligent IT infrastructure built on the pervasive computing paradigm. Furthermore, there is a need to protect investments already made in the existing IT infrastructure (Gupta and Moitra, 2004). Developing E-government system based on the IT infrastructure which has played as an bedrock role. Internet allows access to multiple services, as a foundation to support the digital broadcast systems to apply a global digital network. It is a goverment's responsibility to determine the quality and quantity of the telecommunications networks to handle the new traffic resulting from the use of these new services' level of service quality (Wanga, Caob, Leckiea and Zhang, 2004). Hence, the authors of this research hypothesize that IT infrastructure affects E-Government Adoption positively

*H8: IT Infrastructure will have a positive effect to Critical Factors for E-Government Adoption.*

### 3.4.2. IT Standards:

IT standards are "specifications for hardware and software that are either widely used and accepted or sanctioned by a standard organisation" (Freeman, 2001). Basing on Wakid and Radack (1997), Information Technology Standards refers to the technical rules and the foundation for interconnected systems that work across organisations and geographic locations. IT Standard is one of the factor to define some component of an IT system for many users can use on offerings from multiple sources and multiple vendors (Libicki, 1995). Many standards efforts are begun primarily because of technological reasons (Morell and Stewart, 1996). Moreover, IT creations communicate meaningfully with each other is the reason for the core of the drive behind standar (Libicki, 1995).

Single integrated gateway (one-stop shop) model for adoption of E-government is expected to provide access to its information and services, that requires the goverment public sectors must share information, knowledge, participate positively, and collaborate to provide E-government services. Standards for IT play important role in helping people to manage and use the technology. In many field of technology, standards have a long, useful life, it is so difficult to develop IT standards that are timely and long-lived because of rapid change (Hogan and Radack, 1997). Powerful forces and subject standards-making high uncertainty, such as changing technologies (Morell and Stewart, 1996). IT Standards influence to reduce IT costs of organisations, to facilitate enterprise-wide integration, and to promote greater levels of IT responsiveness (Kayworth and Sambamurthy, 1997). Basing on Wangler, Persson and Soderstrom (2001), a standard can influence in connecting organisational processes and systems, and it also allows a flexible approach in organisational co-operation. Hence we propose the hypothesize:

*H9: IT Standards will have a positive effect to Critical Factors for E-Government Adoption.*

### 3.4.3. National information infrastructure:

As for National Information Infrastructure (NII), Wilson (1997) argues that a NII is the "computerized networks, intelligent terminals and accompanying applications and services people use to access, create, disseminate and utilize digital information". Such infrastructure consists of the physical technologies for example Global Network (GAN), landlines and telecommunication systems. The NII is an extremely important development (Doctor, 1994), and should be accessible to all citizens (Schaefer, 1995). Furthermore, disabled people would be able to access the NII without much inconvenience or expense (Stamoulis and Georgiadis, 2000). Basing on Nambisan and Agarwal (1998), the appearing of Internet, with advancements in telecommunications, has created the number of opportunities for organisations and users to communicate and share information.

IT is an enabler in the public sector. The questions are when and how IT can play a role in helping networks to provide faster, better, and cheaper public E-services. Basing on the General Accounting Office (2001: 1), E-government networks are organisations's network which use ICT for the public of access and delivery of government services. The important key is building up a communications infrastructure to develop E-government system, not only a sufficient distribution of computer technology or social, but also the general applying of telecommunication services is an essential for the attainment of a certain standard of E-government system. The telecommunications fees, for example Internet connection and rental of lines, have an effect in allowing the new capabilities only to finance.

*H10:* *National Information Infrastructure (NII) will have a positive effect to Critical Factors for E-Government Adoption.*

### 3.4.4. Collaboration:

The socially collaborative has the role is that how it can be supported by IT has been of increasing interest to focus on knowledge, sharing and transfer in organisations. Increasing working and collaboration at a distance have maken implementation of groupware technologies (e. g. Lotus Notes) by many organisations including translational corporations operating worldwide (Walsham, 2001). Information sharing for all sites of presence government removes barriers in citizen or business and synchronisation of the relevant information pertaining (Stamoulis and Georgiadis, 2000). Moreover, Caudle et al. (1991) found that the integration of technologies was the most important issue of public sector managers. Integration of functions and department collaboration is necessary to the successful delivery of E-services (Jupp, 2001; Lapre et al, 2001).

Government public sectors are individualised and make decisions by them. However, the important key is an effective communication between departments and public sectors. For this issue, Government must be developed and utilised for internal applications by employees to share information and communication. E-mail is the most success story for improved communications. Bureaucracy and delays have been some consequences for E-government services to the public and toother organisations, be they private or governmental at different levels (Becker, George, Goolsby and Grissom, 1998). The adoption of e-government initiatives necessitates collective efforts from many government public sectors and functional units. E-mail, video conference, discussion forums, use of shared documents, etc., are supported for the efficient and productive collaboration of E-government (Lambrinoudakisa, Gritzalisa, Dridib and Pemul, 2003). Therefore, collaboration is hypothesized to have an indirect impact on E-Government Adoption

*H11:* *Collaboration will have a positive effect to Critical Factors for E-Government Adoption.*

### 3.5. Diffusion of innovation model

According to Rogers (1995), a reason for the very much dwelling interest in the Diffusion of Innovations is because "getting a new idea adopted, even when it has obvious advantages, is very difficult" (Rogers, 1995, p. 1). The main concern of the innovation diffusion research centers on how innovations are adopted as well as the reasons behind innovations is adopted at different rates. Rogers (1995) goes on stating there are four main elements of diffusion innovation, time, communication, and social system. Expanding from the four main elements, Rogers defines diffusion as "the process by which an innovation is communicated through certain channels over time among the members of a social system" (Rogers, 1995, p. 5). The individual's decision on whether to use the technology is based on perception of the technology such as compatibility, relative advantage, image and complexity (Gilbert, D. and Balestrini, P., 2004). Previous study has identified the link between perceived usefulness (relative advantage in innovation diffusion theory) and compatibility (Rogers, 2003). The argument revealed that if the individual perceives an innovation to be inconsistent with his current practice, he would tend to be more uncertain about the expected benefits of the innovation. According to Rogers (2003), the social prestige that the innovation presents to its adopter may be the sole benefit that the adopter receives. From the previous research done by Phang, Li, Sutanto, and Kankanhalli (2005), in the context of Central Provident Fund (CPF) e-Withdrawal service adoption by senior citizens, those senior citizens that adopted the service may impress others that even though they are old in age, they are still able to learn and use up to date technologies which in line with the changes in society. This may enhance their social status and enable them to serve as role models for other senior citizens who have not adopted this e-government service. Based on the revision theoretical framework, the initial hypotheses have to be revised to reflect the actual hypotheses for testing. The re-stated hypotheses are listed as below:

**H12**: *Higher levels of perceived compatibility will be positively related to higher levels of intention to use e-government services.*

**H13**: *Higher levels of perceived relative advantage will be positively related to higher levels of intention to use e-government services.*

**H14**: *Higher levels of perceived image will be positively related to higher levels of intention to use e-government services.*

**H15**: *Higher levels of perceived complexity will be negatively related to higher levels of intention to use e-government services.*

Several testable statements, or hypotheses, can be drawn from the theoretical framework. In present study, the integrated theoretical framework was tested against Vietnam citizens in order to evaluate the factors influencing the intention to use e-government services in Vietnam, and to verify its antecedents. Based on hypotheses used to study the critical factors affecting E-Government adoptions are presented next. The researcher presents the developed model for e-government adoption in term of Vietnam case study. This model consists of the 15 factors and is shown in Figure 2 as the following:

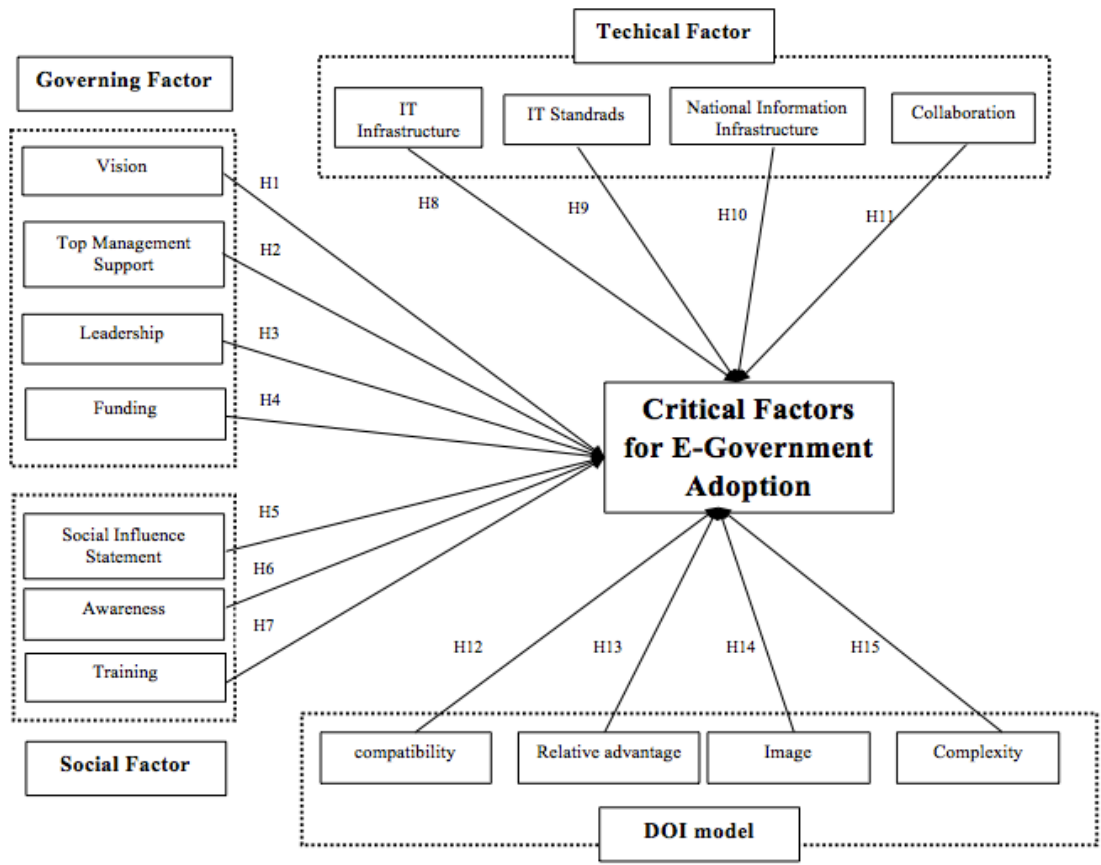

*Figure 2: The theoretical framework.*

## 4. Research method and analysis techniques

We will develop a survey questionnaire for this study. The questionnaire will be designed based on the research conceptual model (see Figure 2). Items will be adapted from prior works on innovation deployment and diffusion related to the concepts this paper advances as discussed in the earlier sections. Responses to the survey questions will be entered on a Five-point Likert-type scale as follows: 1 = Strongly Disagree, 2 = Disagree, 3 = Neutral, 4 = Agree, and 5 = Strongly Agree. The survey questionnaire will include data on participants' profile: sex, age, combined household income, education, job position, family size, and the ethnicity of participants.

We will select the sample of about 450 potential public employee (a quarter is public sector managers) in Vietnam government to test guidelines, theoretical framework and prescriptions provided in the article. The target public employee will be residents in Ho Chi Minh city, Vietnam. We will then take several steps to ensure data validity and reliability. Initially, the questionnaire will be pre- tested with two academics resident in Ho Chi Minh city and two academics resident outside Malacca. The questionnaire will then be revised for any potentially confusing items, before the administering the pilot survey. A pilot survey is aimed at providing an opportunity to objectively measure validity and reliability of the questionnaire (U. Sekaran, 2003). Based on the above recommendations, a pilot study for this research is necessary in developing the survey questionnaire. The pilot study will be conducted using a selected group of 50 residents in Ho Chi Minh city. The suggestions and comments from the pilot study will be evaluated, and those considered relevant will be incorporated into the survey or test design prior to the actual study. We will then use personal questionnaire administration to collect data for this research. To establish the absence of non-response bias, it is desirable to collect data from a set of non-respondents and compare it to data supplied willingly. For a meaningful number of surveys and for all survey items, this method is rarely achievable. A practical preference, that has been argued to provide reliable results, is to compare the mean values of responses for earlier returns with the means from later returns (D. Compeau, 1995). This approach has the capacity to reveal any differences between early and late responders who required prompting. The assumption is that late responders share similarities with non-responders, and if no significant differences exist, the probability is strong that non-response bias does not exist (J. S. Armstrong and T. Overton, 1977). We will conduct tests for all the constructs between first week respondents and those who responded after five weeks, and then determine the differences between the two groups.

Following the response from the online survey, the proposed hypotheses will be tested. SEM based analysis techniques will be used to analysis the data. First, the Confirmatory Factor Analysis (CFA) will be employed to assess the validity of the measurement for the model then the proposed model will be tested using the Structural Equation Modeling (SEM), so that the causal structure of the model can be evaluated. The research will use LISREL 8.7 to analyze the measurement model and the structural model

## 5. Conclusion

This thesis reports on empirical research into the citizen adoption of e-government services in the Vietnam. It is motivated by the problem of the low-level of citizen adoption of e-government services in developing countries such as Vietnam. E-government services cannot improve public service delivery if they are not used by the public. Therefore, the principal objective of this research is to gain a better understanding of the factors that influence the citizens' adoption of e-government services. Identifying such factors will improve the likelihood of increasing the adoption rate of these services by deepening the knowledge about the factors which facilitate, or hinder, the adoption process.

The second motivation of this study is the lack of empirical e-government services adoption research that focuses on the adoption of such services in Vietnam. Therefore, filling this gap in the literature is one of the motivations for conducting this study in a country such as Vietnam, with different cultural and social values. In addition, qualitative research was also employed through the case study, which included semi-structured interviews with the e-government officials in the Ministry of Information and Communication Technologies and the National Information Technology Centre (NITC) – Vietnam. These enabled the researcher to understand, in depth, the factors that influence the adoption of e-government services by the

Vietnam from a managerial perspective. The case study analysis will allow the comparison of the theoretical findings with the actual practice.